\documentclass[lettersize,journal]{IEEEtran}
\IEEEoverridecommandlockouts
\usepackage{cite}
\usepackage{amsmath,amssymb,amsfonts}
\usepackage{graphicx}
\usepackage{textcomp}
\usepackage{xcolor}
\def\BibTeX{{\rm B\kern-.05em{\sc i\kern-.025em b}\kern-.08em
    T\kern-.1667em\lower.7ex\hbox{E}\kern-.125emX}}

\usepackage{pdfpages}
\usepackage{epsfig}
\usepackage{color}
\usepackage{amsxtra}
\usepackage{enumitem}
\usepackage{acro}
\usepackage[T1]{fontenc}
\usepackage{setspace}

\usepackage[cmintegrals]{newtxmath}
\usepackage{graphicx}
\graphicspath{{Figures/}}
\usepackage{url}
\usepackage{cite}
\usepackage{algorithm}
\usepackage{algpseudocode}

\makeatletter
\def\BState{\State\hskip-\ALG@thistlm}
\makeatother

\usepackage{epstopdf}

\usepackage{amsxtra}
\usepackage{amstext}

\usepackage{latexsym}
\usepackage{dsfont} 
\usepackage{color}
\usepackage{multirow}
\usepackage{float}
\usepackage[bookmarks=false]{hyperref}
\usepackage{eucal}
\usepackage{tabularx,colortbl}

\usepackage{lscape}

\usepackage{algorithm}
\usepackage{algpseudocode}

\usepackage{units}
\usepackage{cases}
\usepackage[ subrefformat=parens,labelformat=parens, caption=false, font=footnotesize]{subfig}
\usepackage{mathtools}
\usepackage{pgfplots}
\usepackage{pgf}
\usepackage{tikz}
\pgfplotsset{compat=1.15}

\usepackage{mathtools}

\newcommand{\TT}{\mathsf{T}}
\newcommand{\HH}{\mathsf{H}}

\DeclareAcronym{thz}{
  short = THz,
  long = Terahertz
}

\DeclareAcronym{ris}{
  short = RIS,
  long  = Reconfigurable intelligent surface
}

\DeclareAcronym{lmmse}{
  short = LMMSE,
  long = linear minimum mean square error
}
\DeclareAcronym{ls}{
  short = LS,
  long = least square
}

\DeclareAcronym{simo}{
  short = SIMO,
  long =  single-input multiple-output
}
\DeclareAcronym{BS}{
  short = BS,
  long =  base station
}
\DeclareAcronym{UE}{
  short = UE,
  long =  user equipment
}
\DeclareAcronym{MG}{
  short = MG,
  long = mixture gamma  
}

\DeclareAcronym{MSE}{
  short = MSE,
  long = mean square error  
}

\DeclareAcronym{TDD}{
  short = TDD,
  long = time-division duplex  
}

\DeclareAcronym{PDF}{
  short = PDF,
  long = probability density function  
}

\DeclareAcronym{ML}{
  short = ML,
  long = maximum likelihood 
}

\DeclareAcronym{ULA}{
  short = ULA,
  long = uniform linear array 
}

\DeclareAcronym{los}{
  short = LoS,
  long = line-of-sight 
}

\DeclareAcronym{MC}{
  short = MC,
  long = mutual coupling 
}
\DeclareAcronym{NMSE}{
  short = NMSE,
  long = normalized mean squared error
}

\DeclareAcronym{iid}{
  short = IID,
  long = independent and identically distributed
}

\DeclareAcronym{snr}{
  short = SNR,
  long = signal-to-noise ratio
}

\DeclareAcronym{gbsm}{
  short = GBSM,
  long = geometry-based stochastic model
}

\makeatother

\makeatletter
\def\ps@IEEEtitlepagestyle{%
  \def\@oddfoot{\mycopyrightnotice}%
  \def\@oddhead{\hbox{}\@IEEEheaderstyle\leftmark\hfil\thepage}\relax
  \def\@evenhead{\@IEEEheaderstyle\thepage\hfil\leftmark\hbox{}}\relax
  \def\@evenfoot{}%
}

\def\mycopyrightnotice{%
  \begin{minipage}{\textwidth}
  \centering \scriptsize
    This work has been accepted by the IEEE Communications Letters for publication.  Copyright may be transferred without notice, after which this version may no longer be accessible.
  \end{minipage}
}
\makeatother

\begin{document}
\newcommand{\red}[1]{{\color{red}{#1}}} 
\newcommand{\blue}[1]{{\color{blue}{#1}}} 
\newcommand{\green}[1]{{\color{green}{#1}}} 
\newcommand{\yellow}[1]{{\color{yellow}{#1}}} 
\newcommand{\orange}[1]{{\color{orange}{#1}}} 

\newcommand{\nth}[1]{{#1}{\text{th}}}
\newcommand{\mbf}[1]{\mathbf{#1}}

\newcommand{\Hpow}{{\sf H}}
\newcommand{\Tpow}{{\sf T}}
\newcommand{\Invpow}{{\sf -1}}
\newcommand{\Strpow}{{\sf *}}

\newcommand{\PseuInvpow}{\mathrm{\dagger}}
\newcommand{\npow}{\mathrm{n}}
\newcommand{\Npow}{\mathrm{N}}
\newcommand{\vpow}{\mathrm{v}}
\newcommand{\cpow}{\mathrm{c}}
\newcommand{\ipow}{\mathrm{i}}
\newcommand{\jpow}{\mathrm{j}}
\newcommand{\bpow}{\mathrm{b}}
\newcommand{\rpow}{\mathrm{r}}
\newcommand{\upow}{\mathrm{u}}
\newcommand{\Upow}{\mathrm{U}}
\newcommand{\Bpow}{\mathrm{B}}
\newcommand{\Rpow}{\mathrm{R}}
\newcommand{\tpow}{\mathrm{t}}
\newcommand{\hpow}{\mathrm{h}}
\newcommand{\epow}{\mathrm{e}}
\newcommand{\dpow}{\mathrm{d}}
\newcommand{\Dpow}{\mathrm{D}}
\newcommand{\ppow}{\mathrm{p}}
\newcommand{\kpow}{\mathrm{k}}
\newcommand{\Mpow}{\mathrm{M}}
\newcommand{\mpow}{\mathrm{m}}
\newcommand{\Kpow}{\mathrm{K}}
\newcommand{\Fpow}{\mathrm{F}}
\newcommand{\qpow}{\mathrm{q}}
\newcommand{\abs}[1]{\left|{#1}\right|}
\newcommand{\norm}[1]{\left\|{#1}\right\|}
\newcommand{\vect}[1]{\mathrm{vec}\left(#1\right)}
\newcommand{\supp}[1]{\mathrm{supp}\left(#1\right)}
\newcommand{\trc}[1]{\mathrm{tr}\left(#1\right)}
\newcommand{\diagopr}[1]{\mathrm{diag}\left(#1\right)}
\newcommand{\blkdiagopr}[1]{\mathrm{blkdiag}\left(#1\right)}

\newcommand{\atantwo}{\mathrm{arctan2}}

\newcommand{\lmmseidx}{\mathrm{LMMSE}}
\newcommand{\lsidx}{\mathrm{LS}}
\newcommand{\txidx}{\mathrm{T}}
\newcommand{\rxidx}{\mathrm{R}}
\newcommand{\strmidx}{\mathrm{S}}
\newcommand{\froidx}{\mathrm{F}}
\newcommand{\sysidx}{\mathrm{sys}}
\newcommand{\RFidx}{\mathrm{RF}}
\newcommand{\BBidx}{\mathrm{BB}}
\newcommand{\quantidx}{\mathrm{quant}}
\newcommand{\effidx}{\mathrm{eff}}
\newcommand{\totsupsc}{\mathrm{tot}}
\newcommand{\firstelement}{\mathrm{st}}

\newcommand{\pilotidx}{\mathrm{p}}
\newcommand{\Beamsupsc}{\mathrm{beam}}
\newcommand{\CP}{\mathrm{CP}}
\newcommand{\cent}{\mathrm{cen}}

\newcommand{\coh}{\mathrm{coh}}
\newcommand{\train}{\mathrm{tr}}
\newcommand{\ovs}{\mathrm{ovs}}
\newcommand{\DicRed}{\mathrm{RD}}
\newcommand{\LessCol}{\mathrm{lc}}

\newcommand{\clu}{\mathrm{clu}}
\newcommand{\ray}{\mathrm{ray}}
\newcommand{\LoS}{\mathrm{los}}
\newcommand{\NLoS}{\mathrm{nlos}}
\newcommand{\subb}{\mathrm{sub}}
\newcommand{\GMM}{\mathrm{GMM}}
\newcommand{\svv}{\mathrm{sv}}

\newcommand{\samp}{\mathrm{s}}
\newcommand{\ula}{\mathrm{ULA}}
\newcommand{\upa}{\mathrm{UPA}}
\newcommand{\bmsp}{\mathrm{bmsp}}

\newcommand{\NFsupsc}{\mathrm{NF}}
\newcommand{\FFsupsc}{\mathrm{FF}}
\newcommand{\SWMsupsc}{\mathrm{SWM}}
\newcommand{\HSPMsupsc}{\mathrm{HSPWM}}
\newcommand{\PWMsupsc}{\mathrm{PWM}}

\newcommand{\umarridx}{\mathrm{UMA}}
\newcommand{\saidx}{\mathrm{SA}}
\newcommand{\aeidx}{\mathrm{AE}}

\newcommand{\rotsupsc}{\mathrm{rot}}
\newcommand{\trialsupsc}{\mathrm{trl}}
\newcommand{\maxsupsc}{\mathrm{max}}
\newcommand{\Polsupsc}{\mathrm{pol}}
\newcommand{\DFTsupsc}{\mathrm{dft}}

\newcommand{\allidx}{\mathrm{all}}
\newcommand{\distsupsc}{\mathrm{dist}}
\newcommand{\offsupsc}{\mathrm{offline}}
\newcommand{\onsupsc}{\mathrm{online}}
\newcommand{\estidx}{\mathrm{est}}
\setlength\unitlength{1mm}

\newcommand{\insertfig}[3]{
\begin{figure}[htbp]\begin{center}\begin{picture}(120,90)
\put(0,-5){\includegraphics[width=12cm,height=9cm,clip=]{#1.eps}}\end{picture}\end{center}
\caption{#2}\label{#3}\end{figure}}

\newcommand{\insertxfig}[4]{
\begin{figure}[htbp]
\begin{center}
\leavevmode \centerline{\resizebox{#4\textwidth}{!}{\input
#1.pstex_t}}
\caption{#2} \label{#3}
\end{center}
\end{figure}}

\long\def\comment#1{}



\newfont{\bbb}{msbm10 scaled 700}
\newcommand{\CCC}{\mbox{\bbb C}}

\newfont{\bb}{msbm10 scaled 1100}
\newcommand{\CC}{\mbox{\bb C}}
\newcommand{\PP}{\mbox{\bb P}}
\newcommand{\RR}{\mbox{\bb R}}
\newcommand{\QQ}{\mbox{\bb Q}}
\newcommand{\ZZ}{\mbox{\bb Z}}
\newcommand{\FF}{\mbox{\bb F}}
\newcommand{\GG}{\mbox{\bb G}}
\newcommand{\EE}{\mbox{\bb E}}
\newcommand{\NN}{\mbox{\bb N}}
\newcommand{\KK}{\mbox{\bb K}}


\newcommand{\av}{{\bf a}}
\newcommand{\bv}{{\bf b}}
\newcommand{\cv}{{\bf c}}
\newcommand{\dv}{{\bf d}}
\newcommand{\ev}{{\bf e}}
\newcommand{\fv}{{\bf f}}
\newcommand{\gv}{{\bf g}}
\newcommand{\hv}{{\bf h}}
\newcommand{\iv}{{\bf i}}
\newcommand{\jv}{{\bf j}}
\newcommand{\kv}{{\bf k}}
\newcommand{\lv}{{\bf l}}
\newcommand{\mv}{{\bf m}}
\newcommand{\nv}{{\bf n}}
\newcommand{\ov}{{\bf o}}
\newcommand{\pv}{{\bf p}}
\newcommand{\qv}{{\bf q}}
\newcommand{\rv}{{\bf r}}
\newcommand{\sv}{{\bf s}}
\newcommand{\tv}{{\bf t}}
\newcommand{\uv}{{\bf u}}
\newcommand{\wv}{{\bf w}}
\newcommand{\xv}{{\bf x}}
\newcommand{\yv}{{\bf y}}
\newcommand{\zv}{{\bf z}}
\newcommand{\zerov}{{\bf 0}}
\newcommand{\onev}{{\bf 1}}

\def\u{{\bf u}}


\newcommand{\Am}{{\bf A}}
\newcommand{\Bm}{{\bf B}}
\newcommand{\Cm}{{\bf C}}
\newcommand{\Dm}{{\bf D}}
\newcommand{\Em}{{\bf E}}
\newcommand{\Fm}{{\bf F}}
\newcommand{\Gm}{{\bf G}}
\newcommand{\Hm}{{\bf H}}
\newcommand{\Id}{{\bf I}}
\newcommand{\Jm}{{\bf J}}
\newcommand{\Km}{{\bf K}}
\newcommand{\Lm}{{\bf L}}
\newcommand{\Mm}{{\bf M}}
\newcommand{\Nm}{{\bf N}}
\newcommand{\Om}{{\bf O}}
\newcommand{\Pm}{{\bf P}}
\newcommand{\Qm}{{\bf Q}}
\newcommand{\Rm}{{\bf R}}
\newcommand{\Sm}{{\bf S}}
\newcommand{\Tm}{{\bf T}}
\newcommand{\Um}{{\bf U}}
\newcommand{\Wm}{{\bf W}}
\newcommand{\Vm}{{\bf V}}
\newcommand{\Xm}{{\bf X}}
\newcommand{\Ym}{{\bf Y}}
\newcommand{\Zm}{{\bf Z}}
\newcommand{\Onem}{{\bf 1}}
\newcommand{\Zerom}{{\bf 0}}


\newcommand{\Bc}{{\cal B}}
\newcommand{\Cc}{{\cal C}}
\newcommand{\Dc}{{\cal D}}
\newcommand{\Ec}{{\cal E}}
\newcommand{\Fc}{{\cal F}}
\newcommand{\Gc}{{\cal G}}
\newcommand{\Hc}{{\cal H}}
\newcommand{\Ic}{{\cal I}}
\newcommand{\Jc}{{\cal J}}
\newcommand{\Kc}{{\cal K}}
\newcommand{\Lc}{{\cal L}}
\newcommand{\Mc}{{\cal M}}
\newcommand{\Nc}{{\cal N}}
\newcommand{\Oc}{{\cal O}}
\newcommand{\Pc}{{\cal P}}
\newcommand{\Qc}{{\cal Q}}
\newcommand{\Rc}{{\cal R}}
\newcommand{\Sc}{{\cal S}}
\newcommand{\Tc}{{\cal T}}
\newcommand{\Uc}{{\cal U}}
\newcommand{\Wc}{{\cal W}}
\newcommand{\Vc}{{\cal V}}
\newcommand{\Xc}{{\cal X}}
\newcommand{\Yc}{{\cal Y}}
\newcommand{\Zc}{{\cal Z}}


\newcommand{\alphav}{\hbox{\boldmath$\alpha$}}
\newcommand{\betav}{\hbox{\boldmath$\beta$}}
\newcommand{\gammav}{\hbox{\boldmath$\gamma$}}
\newcommand{\deltav}{\hbox{\boldmath$\delta$}}
\newcommand{\etav}{\hbox{\boldmath$\eta$}}
\newcommand{\lambdav}{\hbox{\boldmath$\lambda$}}
\newcommand{\epsilonv}{\hbox{\boldmath$\epsilon$}}
\newcommand{\nuv}{\hbox{\boldmath$\nu$}}
\newcommand{\muv}{\hbox{\boldmath$\mu$}}
\newcommand{\zetav}{\hbox{\boldmath$\zeta$}}
\newcommand{\phiv}{\hbox{\boldmath$\phi$}}
\newcommand{\psiv}{\hbox{\boldmath$\psi$}}
\newcommand{\thetav}{\hbox{\boldmath$\theta$}}
\newcommand{\tauv}{\hbox{\boldmath$\tau$}}
\newcommand{\omegav}{\hbox{\boldmath$\omega$}}
\newcommand{\xiv}{\hbox{\boldmath$\xi$}}
\newcommand{\sigmav}{\hbox{\boldmath$\sigma$}}
\newcommand{\piv}{\hbox{\boldmath$\pi$}}
\newcommand{\rhov}{\hbox{\boldmath$\rho$}}
\newcommand{\vtv}{\hbox{\boldmath$\vartheta$}}

\newcommand{\Gammam}{\hbox{\boldmath$\Gamma$}}
\newcommand{\Lambdam}{\hbox{\boldmath$\Lambda$}}
\newcommand{\Deltam}{\hbox{\boldmath$\Delta$}}
\newcommand{\Sigmam}{\hbox{\boldmath$\Sigma$}}
\newcommand{\Phim}{\hbox{\boldmath$\Phi$}}
\newcommand{\Pim}{\hbox{\boldmath$\Pi$}}
\newcommand{\Psim}{\hbox{\boldmath$\Psi$}}
\newcommand{\psim}{\hbox{\boldmath$\psi$}}
\newcommand{\chim}{\hbox{\boldmath$\chi$}}
\newcommand{\omegam}{\hbox{\boldmath$\omega$}}
\newcommand{\vphim}{\hbox{\boldmath$\varphi$}}
\newcommand{\Thetam}{\hbox{\boldmath$\Theta$}}
\newcommand{\Omegam}{\hbox{\boldmath$\Omega$}}
\newcommand{\Xim}{\hbox{\boldmath$\Xi$}}


\newcommand{\sinc}{{\hbox{sinc}}}
\newcommand{\diag}{{\hbox{diag}}}
\renewcommand{\det}{{\hbox{det}}}
\newcommand{\trace}{{\hbox{tr}}}
\newcommand{\sign}{{\hbox{sign}}}
\renewcommand{\arg}{{\hbox{arg}}}
\newcommand{\var}{{\hbox{var}}}
\newcommand{\cov}{{\hbox{cov}}}
\newcommand{\SINR}{{\sf sinr}}
\newcommand{\SNR}{{\sf snr}}
\newcommand{\Ei}{{\rm E}_{\rm i}}
\newcommand{\eqdef}{\stackrel{\Delta}{=}}
\newcommand{\defines}{{\,\,\stackrel{\scriptscriptstyle \bigtriangleup}{=}\,\,}}
\newcommand{\<}{\left\langle}
\renewcommand{\>}{\right\rangle}
\newcommand{\herm}{{\sf H}}
\newcommand{\trasp}{{\sf T}}
\renewcommand{\vec}{{\rm vec}}
\newcommand{\calL}{\mbox{${\mathcal L}$}}
\newcommand{\calO}{\mbox{${\mathcal O}$}}

\newcommand{\Afd}{\mbox{$\boldsymbol{\mathcal{A}}$}}
\newcommand{\Bfd}{\mbox{$\boldsymbol{\mathcal{B}}$}}
\newcommand{\Cfd}{\mbox{$\boldsymbol{\mathcal{C}}$}}
\newcommand{\Dfd}{\mbox{$\boldsymbol{\mathcal{D}}$}}
\newcommand{\Efd}{\mbox{$\boldsymbol{\mathcal{E}}$}}
\newcommand{\Ffd}{\mbox{$\boldsymbol{\mathcal{F}}$}}
\newcommand{\Gfd}{\mbox{$\boldsymbol{\mathcal{G}}$}}
\newcommand{\Hfd}{\mbox{$\boldsymbol{\mathcal{H}}$}}
\newcommand{\Ifd}{\mbox{$\boldsymbol{\mathcal{I}}$}}
\newcommand{\Jfd}{\mbox{$\boldsymbol{\mathcal{J}}$}}
\newcommand{\Kfd}{\mbox{$\boldsymbol{\mathcal{K}}$}}
\newcommand{\Lfd}{\mbox{$\boldsymbol{\mathcal{L}}$}}
\newcommand{\Mfd}{\mbox{$\boldsymbol{\mathcal{M}}$}}
\newcommand{\Nfd}{\mbox{$\boldsymbol{\mathcal{N}}$}}
\newcommand{\Ofd}{\mbox{$\boldsymbol{\mathcal{O}}$}}
\newcommand{\Pfd}{\mbox{$\boldsymbol{\mathcal{P}}$}}
\newcommand{\Qfd}{\mbox{$\boldsymbol{\mathcal{Q}}$}}
\newcommand{\Rfd}{\mbox{$\boldsymbol{\mathcal{R}}$}}
\newcommand{\Sfd}{\mbox{$\boldsymbol{\mathcal{S}}$}}
\newcommand{\Tfd}{\mbox{$\boldsymbol{\mathcal{T}}$}}
\newcommand{\Ufd}{\mbox{$\boldsymbol{\mathcal{U}}$}}
\newcommand{\Vfd}{\mbox{$\boldsymbol{\mathcal{V}}$}}
\newcommand{\Wfd}{\mbox{$\boldsymbol{\mathcal{W}}$}}
\newcommand{\Xfd}{\mbox{$\boldsymbol{\mathcal{X}}$}}
\newcommand{\Yfd}{\mbox{$\boldsymbol{\mathcal{Y}}$}}
\newcommand{\Zfd}{\mbox{$\boldsymbol{\mathcal{Z}}$}}

\bstctlcite{IEEEexample:BSTcontrol}

\title{
THz-Band Near-Field RIS Channel Modeling for Linear Channel Estimation
}

\author{
        Ahmad~Dkhan,~\IEEEmembership{Student Member,~IEEE,}
        Simon~Tarboush,
        Hadi Sarieddeen,~\IEEEmembership{Senior Member,~IEEE,}
        
        Ibrahim~Abou-Faycal,~\IEEEmembership{Member,~IEEE,}
        and~Tareq~Y. Al-Naffouri,~\IEEEmembership{Fellow,~IEEE}%
\thanks{Ahmad~Dkhan, Hadi~Sarieddeen, and Ibrahim~Abou-Faycal, are with the Department of Electrical and Computer Engineering (ECE), AUB, Beirut, Lebanon (amd53@mail.aub.edu, hadi.sarieddeen@aub.edu.lb, ia14@aub.edu.lb).
Simon~Tarboush and Tareq Y. Al-Naffouri are with the Department of Computer, Electrical and Mathematical Sciences and Engineering (CEMSE), KAUST, Kingdom of Saudi Arabia (simon.w.tarboush@gmail.com, tareq.alnaffouri@kaust.edu.sa). This work was supported by the AUB University Research Board and the KAUST Office of Sponsored Research (OSR) under Award No. ORFS-CRG12-2024-6478.}
}
\maketitle
\begin{abstract}
Reconfigurable intelligent surface (RIS)-aided terahertz (THz)-band communications are promising enablers for future wireless networks. However, array densification at high frequencies introduces significant challenges in accurate channel modeling and estimation, particularly with THz-specific fading, mutual coupling (MC), spatial correlation, and near-field effects. In this work, we model THz outdoor small-scale fading channels using the mixture gamma (MG) distribution, considering absorption losses, spherical wave propagation, MC, and spatial correlation across large base stations and RISs. We derive the distribution of the cascaded RIS-aided channel and investigate linear channel estimation techniques, analyzing the impact of various channel parameters. Numerical results based on precise THz parameters reveal that accounting for spatial correlation, MC, and near-field modeling substantially enhances estimation accuracy, especially in ultra-massive arrays and short-range scenarios. These results underscore the importance of incorporating these effects for precise, physically consistent channel modeling.
\end{abstract}


\section{Introduction}
\ac{thz}-band communications promise ultra-low latency and terabit-per-second data rates by leveraging the vast spectrum for future wireless networks~\cite{Sarieddeen2021Overview}. Beyond fully autonomous vehicles, extended reality, and massive connectivity, THz systems also enable holographic communications and novel sensing and localization capabilities, making this band a key enabler for sixth-generation networks and beyond \cite{Sarieddeen2020Next}. 

However, the practical implementation of \ac{thz} networks faces significant challenges due to severe spreading and molecular absorption losses~\cite{Tarboush2021Teramimo}. Furthermore, propagation at \ac{thz} frequencies is more susceptible to blockage and, in many cases, relies on \ac{los} paths, significantly reducing link availability. \acp{ris} with massive elements provide a viable solution to mitigate {such} propagation challenges by dynamically controlling signal reflections and offering alternative communication paths~\cite{Basar2019Wireless}.

Integrating \ac{ris} into \ac{thz} communications has gained attention~\cite{Matos2024Reconfigurable}. However, accurate channel modeling must account for \ac{thz}-specific propagation, massive arrays, and surfaces. The short wavelength and large apertures result in a significant near-field region~\cite{Tarboush2024Cross}, requiring spherical wave modeling rather than the planar wave model. Additionally, dense array configurations increase spatial correlation and \ac{MC} between elements and must be considered~\cite{KolomvakisExploiting,Gong2024Near}. Conventional models typically assume correlated Rayleigh fading with far-field spatial correlation matrices~\cite{Bjornson2021Rayleigh} and, more recently, near-field scenarios~\cite{Delbari2024Far}. The authors in \cite{Li2023Performance} have derived closed-form expressions for outage probability and ergodic channel capacity that account for channel estimation errors and \ac{ris} phase shift noise by using the moment-matching method and approximate distributions. However, the previous studies and assumptions are not suitable for \ac{thz} channels with limited non-\ac{los} paths and massive arrays and neglect the \ac{MC} effect. Furthermore, several works have shown that ignoring \ac{MC} will drastically degrade the estimation performance ~\cite{KolomvakisExploiting,Zheng2024Mutual}.

The \ac{MG} distribution has been applied to study the physical layer security of \ac{ris}-aided systems~\cite{Boulogeorgos2023Ergodic} because of its flexibility in modeling various well-known distributions. Empirical studies further confirm the \ac{MG} distribution's superior fit for small-scale fading in outdoor \ac{thz} links~\cite{Papasotiriou2023Outdoor}, with~\cite{Jemaa2024Performance} analyzing its performance in point-to-point \ac{thz} links. Furthermore, assuming \ac{iid} Nakagami-\textit{m} fading, \ac{MG} distribution has been used to derive a tractable cascaded channel model \cite{Li2024Analysis}. However, \cite{Li2024Analysis} assumes a single-antenna \ac{BS} and neglects \ac{MC} and spatial correlation, underscoring the need for further analysis of correlated \ac{MG} channels.

One of the key challenges in \ac{ris}-aided links is channel estimation. Foundational works~\cite{You2020Intelligent, Jensen2020Optimal} have analyzed the \ac{ls}~\cite{Jensen2020Optimal} and the \ac{lmmse} estimators~\cite{You2020Intelligent}. However, these linear channel estimation methods must be adapted for \ac{thz} bands. 

In this work, we propose a hybrid channel modeling methodology for \ac{thz}-band \ac{ris}-aided links. Specifically, our approach combines \ac{gbsm} and correlation-based models by adopting Kronecker-based correlated \ac{MG} distributions, and the correlation matrices, suitable for large arrays and surfaces, account for spatial correlation, \ac{MC}, near field propagation, large-scale \ac{thz}-specific losses, and capture the sparsity in delay- and spatial-domains by employing the multi-ring approach. In such accurate modeling, the \ac{ls} and \ac{lmmse} estimators have been evaluated. Our results demonstrate that accounting for near-field spatial correlation and \ac{MC} improves estimation accuracy by several dBs compared to models that neglect such effects.

\section{System Model and Problem Formulation}\label{sec:sys_mod}
\begin{figure}[t]
    \centering
    \includegraphics[width=0.85\linewidth]{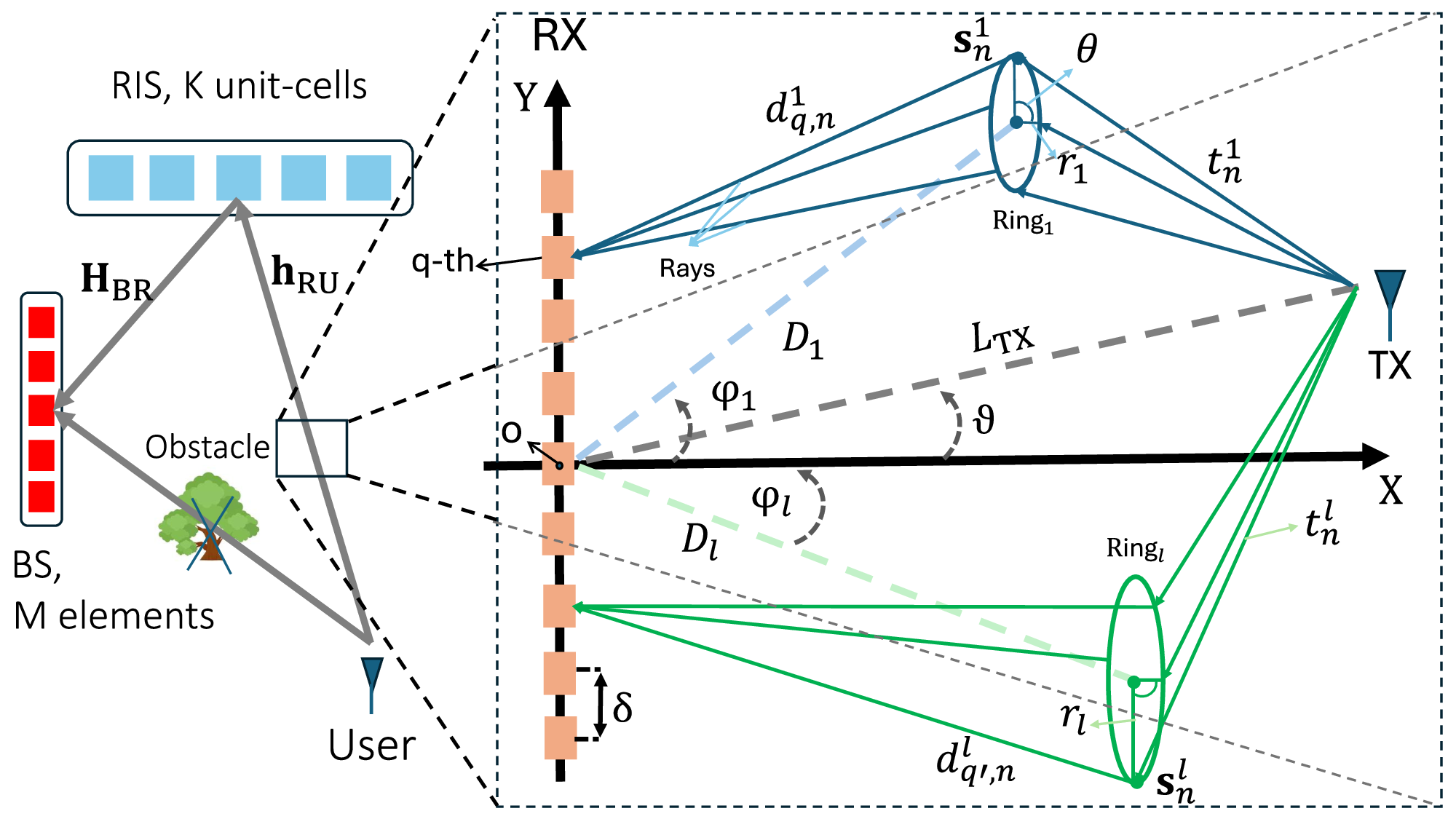}
    \caption{Illustration of an uplink SIMO \ac{ris}-aided \ac{thz} communication system.}
    \label{fig:image1}
    \vspace{-5mm}
\end{figure}

We consider the uplink of a \ac{TDD} \ac{ris}-aided \ac{simo} \ac{thz} communication system, where a single-antenna \ac{UE} communicates with a \ac{BS} with \( M \) antennas, through a nearly-passive \( K \)-unit-cell \ac{ris}. To focus on the \ac{ris}-related problem, we assume that the direct \ac{UE}-\ac{BS} channel is blocked, as shown in Fig.~\ref{fig:image1}. This model employs the Cartesian coordinate system, where the \ac{ris} adopts a \ac{ULA} along the Y-axis. The $\nth{k}$ unit-cell location is \( \pv_k^\Rpow = [ 0, k\delta_\Rpow, 0]^\TT \), where \( -\left\lceil (K-1)/2 \right\rceil \leq k \leq \left\lfloor (K-1)/2 \right\rfloor \), and \( \delta_\Rpow \) is the \ac{ris} inter-element spacing. Similarly, the $\nth{m}$ antenna-element location is \( \pv_m^\Bpow = [ 0, m\delta_\Bpow, 0]^\TT \), where \( -\left\lceil (M-1)/2 \right\rceil \leq m \leq \left\lfloor (M-1)/2 \right\rfloor \), and \( \delta_\Bpow \) is the \ac{BS} inter-antenna spacing. The \ac{UE} location is \( \mathbf{p}^\Upow = L_\Upow[ \cos\vartheta_\Upow, \sin\vartheta_\Upow, 0]^\TT \), where \( L_\Upow \) denotes the distance between the \ac{UE} to the origin, and the angle \( \vartheta_\Upow \in \left[-\frac{\pi}{2}, \frac{\pi}{2}\right] \) is measured relative to the positive X-axis.

The uplink channel-estimation training procedure involves adjusting the phase shifts of the \ac{ris} elements. At the \( \ell \)th training step, with a training duration, $T_\ppow=K$, the received signal at the \ac{BS} is expressed as~\cite{Jensen2020Optimal,You2020Intelligent,Pan2022Overview}
\begin{equation}
\label{MRIs}
\yv_\ell = \Hm_{\Bpow\Rpow}\diagopr{\phiv_\ell} \hv_{\Rpow\Upow} x_\ell+ \nv_\ell=\Cm\boldsymbol{\phi}_\tpow x_\ell + \nv_\ell, 
\end{equation}
where \(\yv_\ell \!\in\! \CC^M\) is the \ac{BS} received signal, \(\Hm_{\Bpow\Rpow} \!\in\! \CC^{M \times K}\) and \(\hv_{\Rpow\Upow} \!\in\! \CC^K\) are the \ac{ris}-\ac{BS} and \ac{UE}-\ac{ris} \ac{thz} correlated channels (further details in Sec.~\ref{sec:thz_ch_mol} and Sec.~\ref{sec:corr_model}), respectively, \(\phiv_\ell = [\phi_{\ell,1}, \cdots,\phi_{\ell,K}]^\TT\!=\![e^{j\nu_{\ell,1}}, \cdots, e^{j\nu_{\ell,K}}]^\TT \!\in\! \CC^K\) is the \ac{ris} phase shift vector with \(\nu_{\ell,k} \!\in\! [0, 2\pi)\), \(\ell \!\in\! \{1, \ldots, T_\ppow\}\), and \(k \!\in\! \{ 1, \ldots, K\}\). The matrix $\diagopr{\phiv_\ell}$ is a diagonal matrix with diagonal entries $\{\!\phi_{\ell,1}, \cdots,\phi_{\ell,K}\!\}$. Additionally, \(x_\ell\) is the \ac{UE} transmitted pilot symbol.
The cascaded channel is defined as \( \Cm= [\cv_1, \cv_2, \cdots, \cv_K] \triangleq \Hm_{\Bpow\Rpow}\diagopr{\hv_{\Rpow\Upow}} \in \CC^{M \times K} \), and \(\nv_\ell \sim \mathcal{CN}(0, \sigma^2 \mathbf{I}_M)\) is the additive white Gaussian noise. By stacking the training time slots, the $\yv_\ell$ vectors are combined into a compact linear measurement model \cite[eq. (8)]{Jensen2020Optimal},
\begin{equation}
\yv =  \Qm\cv + \nv, \label{meq}
\end{equation}
where 
\(\yv = [\yv_1, \dots, \yv_{T_\ppow}]^\TT\), \(\cv = [\cv_1, \dots, \cv_K]^\TT\), and \(\nv = [\nv_1, \dots, \nv_{T_\ppow}]^\TT\). represent observation, unknown, and noise vectors, respectively, with \(\nv \sim \mathcal{CN}(0, \sigma^2 \mathbf{I}_{M T_\ppow})\). Additionally ~\cite{Jensen2020Optimal},
\(\Qm = \Xm \Psim, \label{Q}\) with $\Xm =\diagopr{[x_1\onev_M,\dots,x_{T_\ppow}\onev_M]} \in \mathbb{R}^{T_\ppow M \times T_\ppow M}$ and $\Psim = \Phim \otimes \textbf{I}_M \in \mathbb{C}^{T_\ppow M \times K M}$. The discrete Fourier transform (DFT)-based phase shift matrix,  
\begin{equation}
\label{fi mat}
\Phim = \begin{bmatrix} \phi_{1,1} & \cdots & \phi_{1,K} \\ \vdots & \ddots & \vdots \\ \phi_{T_\ppow,1} & \cdots & \phi_{T_\ppow,K} \end{bmatrix}, \quad
\phi_{\ell,k}  = e^{-j2\pi(\ell-1)(k-1)/T_\ppow},
\end{equation}
minimizes the estimation variance by optimally configuring \ac{ris} phase shifts to enhance signal power and reduce the mean squared error, as outlined in~\cite{You2020Intelligent, Pan2022Overview}.
\section{Proposed Channel Model}
\label{sec:ch_mod}
\subsection{THz Channel Model}
\label{sec:thz_ch_mol}
\subsubsection{The Mixture Gamma Distribution}
\label{sec:mg_thz}
Accurate modeling of \ac{thz} channel propagation is crucial due to the distinct characteristics of this band~\cite{Tarboush2021Teramimo}. The \ac{MG} distribution is well-suited for outdoor \ac{thz} small-scale fading by combining multiple Gamma components with varying parameters~\cite{Papasotiriou2023Outdoor}. The inherent non-negative Gamma distribution makes \ac{MG} models robust against outliers in high-correlation scenarios, suitable for \ac{thz}-band analysis~\cite{Papasotiriou2023Outdoor}. Additionally, \ac{MG} models can adaptively approximate various distributions, including Rayleigh, Rice, Nakagami-\textit{m}, and Gamma~\cite[Table II]{Li2024Analysis}.
Let \(X\) be an \ac{MG}-distributed random variable of \ac{PDF} $f_X(x)\sim\mathcal{MG}(w_j,\alpha_j,\beta_j)$~\cite{Jemaa2024Performance,Papasotiriou2023Outdoor,Li2024Analysis},
\begin{equation}
f_X(x)\!=\!\!\sum_{j=1}^{J}\!\frac{w_j \beta_{j}^{\alpha_j}}{\Gamma(\alpha_j)} x^{\alpha_j - 1} e^{-\beta_j x}\!=\!\!\sum_{j=1}^{J}\!\varepsilon_j x^{\alpha_j - 1} e^{-\beta_j x}, \, x \geq 0,\label{eq:MG}
\end{equation}
where \(J\) is the number of gamma components, \(\alpha_j\) and \(\beta_j\) are the shape and rate of each component, and \(w_j\) is a weight satisfying $\sum_{j=1}^J w_j \!=\! 1$. We define \(\varepsilon_j = \frac{w_j \beta_{j}^{\alpha_j}}{\Gamma(\alpha_j)}\) for ease of notation. The mean and variance of $X$ in~\eqref{eq:MG} are, respectively,
\begin{equation}
\label{eq:MG_mean_var}
\EE[X]\!=\!\sum_{j=1}^{J}\!w_j \frac{\alpha_j}{\beta_j},\!\ \text{and} \  
\sigma_{X}^2\!=\!\sum_{j=1}^{J}\!w_j\!\left(\!\frac{\alpha_j}{\beta_j^2}\!+\!\left( \frac{\alpha_j}{\beta_j} - \EE[X] \right)^2 \right). 
\end{equation}
The variance is influenced by the shape and rate parameters, where $\alpha$ reflects scattering dispersion around the average power, and lower $\beta$ values yield greater variance.

\subsubsection{THz-Specific Channel Characteristics}
\label{sec:thz_charac}
We model an \ac{ris}-aided \ac{thz} channel in the presence of near-field spatial correlation caused by large \ac{BS} and \ac{ris} arrays. For the multiplicative cascaded channel in~\eqref{MRIs}, we first model a \ac{simo} channel using the generalized non-uniform spherical wave (NUSW) model, representing the \ac{UE}-\ac{ris} channel in Fig.~\ref{fig:image1}. We adopt the \ac{ULA}-based model from~\cite{Dong2022Near},
where the small-scale fading is modeled as \ac{MG} distribution~\eqref{eq:MG} and incorporating both spreading and molecular absorption loss. The complex channel coefficient $h_q$ between a transmitter and an $q$th receive array element can be expressed as~\cite{Dong2022Near,Tarboush2021Teramimo}
\begin{equation}
\label{eq:ch_hq_all}
h_{q} =  \sum_{l=1}^{L} \sum_{n=1}^{N_l} \frac{\lambda \varrho_n^l}{4 \pi t_{n}^l d_{q,n}^l} e^{-\frac{K_{\mathrm{a}}}{2} (t_{n}^l + d_{q,n}^l)} e^{-j \frac{2\pi}{\lambda}(t_{n}^l + d_{q,n}^l) + j \psi_{n}^l}, 
\end{equation}
where $L$ and $N_l$ denote the number of dominant spatial clusters (a ring in Fig. \ref{fig:image1} represents each cluster) and the number of scatterers or reflectors in the $l$-th ring (resulting in multipath components with different delays), respectively. Typically, $L$ and $N_l$ are small due to the use of directional massive arrays and severe propagation losses at \ac{thz}-band~\cite{Tarboush2021Teramimo,Tarboush2024Cross}, \(\lambda=\frac{c_0}{f_c}\) is the signal wavelength, $c_0$ is the speed of light, and $f_c$ is the center frequency. The parameter, \(\varrho_n^l\), denotes the $n$th path reflection coefficient magnitude of the $l$th cluster (details in~\cite{Tarboush2024Cross}), \( t_n^l \) is the distance between the transmitter and the $n$th scatterer or reflector of the $l$th cluster, and \( d_{q,n}^l \) accounts for the distance between the $n$th scatterer/reflector associated with the $l$th cluster and the $q$th receiver element. The term, \( \frac{\lambda \varrho_n^l }{4 \pi t_{n}^l d_{q,n}^l} \), is the spreading loss of $n$th path in the $l$th cluster, while the molecular absorption loss is defined as \(e^{-\frac{K_{\mathrm{a}}}{2} (t_{n}^l + d_{q,n}^l)}\) where \( K_{\mathrm{a}} \) is the molecular absorption coefficient (details in~\cite{Tarboush2021Teramimo}). \( \psi_n^l \) denotes the phase shift introduced by the $n$th path within the $l$th cluster, modeled as a random variable uniformly distributed over \( [-\pi, \pi) \) \cite{Dong2022Near}.
By applying some mathematical manipulations, \( h_{q} \) of \eqref{eq:ch_hq_all} can be written as
\begin{equation}
\label{eq:ch_hq_step2}
h_q\!\!=\!\!\!\sum_{l=1}^{L}\!\sum_{n=1}^{N_l}\!\frac{\lambda\!\varrho_n^l e^{-\frac{K_{\mathrm{a}}(\!t_{n}^l+\!d_{n}^l\!)}{2}}}{4\pi t_{n}^ld_{n}^l}\!\frac{d_{n}^l}{d_{q,n}^l} e^{\!-\frac{K_{\mathrm{a}}(\!d_{q,n}^l\!- d_{n}^l\!)}{2}}\!e^{-j \frac{2\pi}{\lambda}(\!t_{n}^l\!+\!d_{q,n}^l\!)+j\!\psi_n^l}\!.\!
\end{equation}
To simplify~\eqref{eq:ch_hq_step2}, we denote \(\frac{\lambda \varrho_n^l e^{-\frac{1}{2} K_{\mathrm{a}}(t_{n}^l + d_{n}^l)}}{4 \pi t_{n}^l d_{n}^l} = \sqrt{\frac{\Omega_l}{N_l}} \tilde{h}_{n}\) where \( \Omega_l = \EE[|h_0|^2] \) is the average received power from the $l$th cluster at the reference array element \( q\!=\!0 \), where $\sum_{l=1}^{L} \Omega_l=\Omega$. \( \tilde{h}_{n} \) is a random variable representing the signal magnitude at \( q\!=\!0 \) contributed by the \( n \)th scatterer/reflector. Assuming that the channel power fraction contributed by cluster \( l \) satisfies \( \sum_{l=1}^L \epsilon_l = 1 \) with \( \epsilon_l > 0 \), the channel $h_q$ is 
\begin{equation}
\label{eq:ch_hq_step3}
h_q\!=\!\sqrt{\Omega}\!\sum_{l=1}^{L}\!\sqrt{\frac{\epsilon_l}{N_l}}\!\sum_{n=1}^{N_l}\!\tilde{h}_{n}\!\frac{d_{n}^l}{d_{q,n}^l}\!e^{-\frac{K_{\mathrm{a}}(d_{q,n}^l\!-\!d_{n}^l)}{2}} e^{-j\!\frac{2\pi}{\lambda}(t_{n}^l\!+\!d_{q,n}^l)\!+\!j\!\psi_n^l}.
\end{equation}
To this end, our model captures the \( l \)th spatial cluster and is centered around a distinct angle \( \varphi_l \), comprising multiple rays with different time delays, thereby capturing the \ac{thz} channel characteristics across the delay and spatial domains.
In conventional channel models, such as under assumptions in~\cite{Dong2022Near}, $h_{q}$ follows a complex Gaussian distribution, motivated by the expected large number of non-LoS paths and rich scatterer environments at low-frequency bands. However, in the \ac{thz} band, the number of non-LoS paths is limited~\cite{Tarboush2021Teramimo}. Thus, we model $h_{q}$ following the \ac{MG} distribution of~\eqref{eq:MG} to capture the magnitude of the small-scale fading. Let $\hv \!=\! [h_1, \dots, h_Q]^\TT \!\in\! \mathbb{C}^{Q \times 1}$ denotes the vector of channel coefficients, $h_q$, of~\eqref{eq:ch_hq_step3}, and $\dv^l_n \!=\! [d^l_{1,n}, \dots, d^l_{Q,n}]^\TT$ denotes the corresponding vector of distances, $d^l_{q,n}$, associated with the $n$th scatterer from the $l$th ring. Then, the channel $\hv$ is defined
\begin{equation}
\label{eq:ch_hq_vect}
\hv\!\!=\!\!\sqrt{\Omega}\!\sum_{l=1}^{L}\!\big(\!\frac{\epsilon_l}{N_l}\!\big)\!^{\frac{1}{2}}\!\!\sum_{n=1}^{N_l}\!\tilde{h}_{n} d_n^l e^{-\!j \frac{2\pi}{\lambda} t_n^l\!+\frac{K_\mathrm{a}d_n^l}{2}\!+j\!\psi_n^l}\!\left(\!\frac{1}{\dv_n^l}\!\odot\!e^{-\!j\!\frac{2\pi}{\lambda} \dv_n^l \!- \!\frac{K_{\mathrm{a}} \dv_n^l}{2}}\!\!\right)\!.\!\!\!
\vspace{-8mm}
\end{equation}
\subsection{Near- and Far-field Spatial Correlation}
\label{sec:corr_model}
\begin{figure*}[!htb]
\vspace{-5mm}
\normalsize
\begin{align}
\label{eq:RNF_sum}
    \Rm_{\text{NF}}(q, p)&\!=\!\EE[\!h_q h_p^\HH]\!=\!\Omega\!\sum_{l=1}^{L}\frac{\epsilon_l}{N_l}\!\sum_{n=1}^{N_l}\!\sum_{v=1}^{N_l}\frac{\!d_n^l d_{v}^l}{d_{n,q}^l d_{v,p}^l} \EE\left[\!e^{j\!\psi_n^l\!-\!j\!\psi_{v}^l}\!\right]e^{-j \frac{2\pi}{\lambda} (t_n^l\!+\!d_{q,n}^l-t_{v}^l-\!d_{p,v}^l)} e^{-\frac{K_{\mathrm{a}}(d_{q,n}^l - d_{n}^l + d_{p,v}^l - d_{v}^l)}{2}} \EE[\tilde{h}_n \tilde{h}_{v}],\notag
    \\& = \Omega \sum_{l=1}^{L}\frac{\epsilon_l}{N_l} \sum_{n=1}^{N_l} \EE[\tilde{h}_{n}^2] \frac{d_n^{l^2}}{d_{q,n}^l d_{p,n}^l} e^{-j \frac{2\pi}{\lambda}(d_{q,n}^l - d_{p,n}^l)}  e^{-\frac{1}{2} K_{\mathrm{a}}( d_{q,n}^l + d_{p,n}^l-2d_n^l)}.
\end{align}
\hrulefill
\vspace*{0pt}
\vspace{-5mm}
\end{figure*}
We explore the spatial correlation matrix using the near and far-field models. We express the $(q,p)$ element of the correlation matrix, $\Rm_{\text{NF}}$, using~\eqref{eq:ch_hq_step3}, as expressed in \eqref{eq:RNF_sum} at the top of next page, utilizing \(\EE\left[ e^{j \psi_n^l - j \psi_v^l} \right]\!=\! \delta(n - v)\) since $\psi_n^l$ are \ac{iid} uniformly distributed random variables over $[-\pi, \pi)$.

Thus, similar to~\cite{Dong2022Near}, we may express \( \EE[\tilde{h}_{n}^2]/N_l \!=\! p_l(\sv^l_n) \, d\sv\), where \( p_l(\sv) \) is the~\ac{PDF} of the scatterer location of the $l $th path and \( \sv^l_n \in \Sm^{l}\) is the location of the $n$th scatterer point from the cluster $l$. However, the multi-ring approach provides a more refined representation of the scatterers/reflectors in the near field, we can re-express this in terms of the angular parameter as \(\EE[\tilde{h}_{n}^2]/N_l = p_l(\theta^l_n) \, d\theta,\) where \( \theta \in [-\pi, \pi) \) and \( p_l(\theta) \) represents the~\ac{PDF} of the scatterer's angular position. Consequently, the near-field spatial correlation is
\begin{align}
\Rm_{\text{NF}}(q,p) &= \Omega \sum_{l=1}^{L}\epsilon_l\int_{-\pi}^{\pi} \frac{d^l(\theta)^2}{d_q^l(\theta) d_{p}^l(\theta)} e^{-j\frac{2\pi}{\lambda}(d_q^l(\theta) - d_{p}^l(\theta))} \notag \\
& \quad \times e^{-\frac{1}{2} K_{\mathrm{a}}( d_q^l(\theta) + d_{p}^l(\theta))- 2d^l(\theta)} 
p_l(\theta) \, d\theta.
\end{align}
The scatterer location can be parameterized using the ring center and radius \(r_l\). The distance between the scatterer and the \( q \)th antenna element is expressed by~\cite{Dong2022Near} 
\begin{equation}
d_q^l\!(\!\theta\!)\!=\!\sqrt{(\!D_l\!\cos(\!\varphi_l\!)\!+\!r_l\!\cos(\!\theta)\!)\!^2\!+\!(\!D_l\!\sin\!(\!\varphi_l\!)\!+\!r_l\!\sin(\!\theta\!)\!-\!q\delta\!)\!^2},
\label{d_theta}
\end{equation}

where \( \varphi_l \in [-\pi/2, \pi/2) \) is the angle between the vector connecting the center of the ring to the center of the \ac{ULA} (distance of \(D_l\)) and the positive X-axis, and  \( \delta \) is the inter-element spacing. To model the angular spread of scatterers, we consider the von Mises \ac{PDF} \cite{Bjornson2021Rayleigh, Dong2022Near},
\begin{equation}
p_l(\theta) = \frac{e^{\kappa_l \cos(\theta - \mu_l)}}{2\pi I_0(\kappa_l)}, \quad \theta \in [-\pi, \pi). 
\label{von}
\end{equation}
where \( I_0(\cdot) \) is the zero-order Bessel function of the first kind, \( \kappa_l \geq 0 \) determines the concentration of the distribution, and \( \mu_l \) corresponds to the mean of the \( l \)th spatial cluster, representing the angle around which the rays in the cluster are distributed. 

In the far-field scenario, the spatial correlation function is
\begin{align}
\Rm_{\text{FF}}(q, p) &=\Omega \sum_{l=1}^{L}\epsilon_l \int_{-\pi}^{\pi} e^{-j \frac{2\pi}{\lambda} (q - p) \delta 
\frac{D_l \sin (\varphi_l) + r_l \sin (\theta)}{\sqrt{D_l^2 + r_l^2 + 2D_l r_l \cos(\varphi_l - \theta)}}} \nonumber \\
& \quad \times e^{-\frac{1}{2} K_{\mathrm{a}}(  d_q^l(\theta) + d_{p}^l(\theta))-2 d^l(\theta) } p_l(\theta) \, d\theta,
\end{align}
where the term \( d_q^l(\!\theta)\!+\!d_{p}^l(\!\theta)\) is approximated as
\begin{align}
    d_q^l(\theta)+d_{p}^l(\theta)&\approx2\sqrt{D_l^2+r_l^2+2D_l r_l\cos(\varphi_l-\theta)}\notag\\&-\frac{(q-p)\delta\left(D_l\sin(\varphi_l)+r_l\sin(\theta)\right)}{\sqrt{D_l^2+r_l^2+2D_lr_l\cos(\varphi_l-\theta)}}.
\label{Taylor_exp}
\end{align}
Hence, far-field spatial correlation is expressed as
\begin{equation}
    \Rm_{\text{FF}}\!(\!q,\!p)\!=\!\Omega \sum_{l=1}^{L}\!\epsilon_l\!\int_{-\pi}^{\pi}\!e^{\frac{\left(\!-j\!\frac{2\pi}{\lambda}\!(\!q\!-\!p\!)\!+\!\frac{1}{2}\!K_{\mathrm{a}}\!(\!q\!+ p\!)\!\right) \delta (D_l\!\sin(\!\varphi_l)\!+\!r\!\sin(\!\theta\!))}{\sqrt{D_l^2\!+\!r_l^2\!+\!2D_l r_l\!\cos(\varphi_l - \theta)}}} p_l(\!\theta\!)\!\, d\!\theta.\notag
\end{equation}
To this end, according to the proposed Kronecker channel model, we can express the channel \(\hv\) of \eqref{eq:ch_hq_vect} as
\begin{equation}
\label{eq:h_corr_mg}
    \hv = \Rm_{\Xi}^{1/2} \tilde{\hv},
\end{equation}
with $\Rm_{\Xi}^{1/2}$ is the spatial correlation matrix $\Big(\Xi\!\in\!\{\mathrm{FF},\!\mathrm{NF}\}\Big)$, and \(\tilde{\hv}\), a vector of uncorrelated \ac{MG}-distributed random variables. 
\vspace{-3mm}
\subsection{RIS-aided Cascaded Channel Model}\label{sec:casris_model}
To analyze the cascaded channel in \eqref{meq}, the first subchannel, \(\hv_{\Rpow\Upow}\), can be expressed using~\eqref{eq:h_corr_mg} as
\begin{equation}
    \hv_{\Rpow\Upow} = \Rm_{\Rpow \Upow}^{1/2} \tilde{\hv}_{\Rpow\Upow},
\end{equation}
where \(\tilde{\hv}_{\Rpow\Upow}\) is \ac{iid} and \ac{MG}-distributed, \(\mathcal{MG}( w_{u},\alpha_{u}, \beta_{u})\). Similarly, the \ac{ris}-\ac{BS} channel, \(\Hm_{\Bpow\Rpow}\), is derived as
\begin{align}
\Hm_{\Bpow\Rpow} &=\sqrt{\Omega_{\Bpow\Rpow}} \sum_{l=1}^{L'}\!\sqrt{\frac{\epsilon'_l}{N_l'}}\!\sum_{n'=1}^{N_l'}\!\tilde{h}_{n'} e^{j\!\psi_{n'}} \left(\!\frac{d_{n'}^{l'}}{\dv_{n'}^{l'}}\!\odot\!e^{-j\!\frac{2\pi}{\lambda}\!\dv_{n'}^{l'} - \frac{1}{2}\!K_{\mathrm{a}}(\!\dv_{n'}^{l'}\!-\!d_{n'}^{l'}\!)}\!\right) \nonumber \\&\quad \times \left( \frac{t_{n'}^{l'}}{\tv_{n'}^{l'}} \odot e^{-j \frac{2\pi}{\lambda} \tv_{n'}^{l'} - \frac{1}{2} K_{\mathrm{a}}(\tv_{n'}^{l'} - t_{n'}^{l'})} \right)^\TT,
\vspace{-3mm}
\end{align}
where $\tv_{n'}^{l'} \!=\! [t_{n',1}^{l'}, \dots, t_{n',K}^{l'}]^\TT$ is the distances for scatterer $n'$ from path $l'$, while $\dv_{n'}^{l'} = [d_{1,n'}^{l'}, \dots, d_{M,n'}^{l'}]^\TT$ encapsulates the distances $d_{m,n'}^{l'}$ corresponding to the set of the scatterers in $\Sm^{l'}_{\Bpow \Rpow}$ within the $l'$th cluster. Similarly,
\begin{equation}
    \Hm_{\Bpow\Rpow} = \Rm_{\Bpow \Rpow}^{1/2} \tilde{\Hm}_{\Bpow\Rpow} \Rm_{\Rpow \Bpow}^{1/2},
    \label{H_BR}
\end{equation}
where \(\tilde{\Hm}_{\Bpow\Rpow}\) is \ac{iid} and follows \(\mathcal{MG}(w_{b},\alpha_{b}, \beta_{b})\). Each element of the cascaded channel is a product of two independent \ac{MG} distributions. Using \cite[Theorem 1]{Li2024Analysis}, this product is also \ac{MG}-distributed, and the \ac{PDF} of the cascaded channel is
\begin{equation}
    g(c_{m,k}) =  \sum_{\hat{\Sfd}} \hat{\varepsilon} \cdot c^{\hat{\alpha}-1} \cdot e^{-c \cdot \hat{\beta}},
\end{equation}
where the summation range $\hat{\Sfd}$ and parameters are defined as
\begin{align}
        \hat{\Sfd}\!&=\!\{1\!\leq\!u\!\leq\!U,~ 1\!\leq b\!\leq B,~ 1\!\leq\!a\!\leq A\},\,\hat{\alpha}\!=\!\alpha_{u},\,\hat{\beta}\!=\!\frac{\beta_{u}\beta_{b}}{t_{a}}, \notag\\
        \hat{\varepsilon} &= \left( \frac{w_{{u}} \beta_{{u}}^{\alpha_{{u}}}}{\Gamma\left(\alpha_{{u}}\right)} \right) \cdot \left( \frac{w_{{b}} \beta_{{b}}^{\alpha_{{u}}}}{\Gamma\left(\alpha_{{b}}\right)} \right) \cdot \varpi_{a} t_{a}^{-\alpha_{u} + \alpha_{{b}} - 1}.
\end{align}
Here, $B$ and $U$ represent the number of components of MG parameters from \(H_{\Bpow\Rpow}\) and \(h_{\Rpow\Upow}\), respectively, \(t_{a}\) is the $a$th root of the Laguerre polynomial, \(L_n(t)\), and \(\varpi_{a}\) is the $a$th weight of the Gaussian-Laguerre quadrature. The Gaussian-Laguerre quadrature is expressed as \(\int_0^\infty e^{-t} f(t) dt \approx \sum_{a=1}^{A} \varpi_{a} f(t_a)\), where \(\varpi_{a} = \frac{t_a}{(n+1)^2 L_{n+1}(t_a)^2}\).
The weight, \( \hat{w} \), is expressed in terms of \( \hat{\varepsilon} \) as  \( \hat{w} = \hat{\varepsilon}\frac{\Gamma(\alpha_{u}) t_a^{\alpha_{u}}}{\beta_{u}^{\alpha_{u}} \beta_{b}^{\alpha_{u}}}.\) To determine the variance of the MG element, \( c_{m, k} \), we apply the formula in \eqref{eq:MG_mean_var}, resulting in
\begin{align}
\sigma_{c_{m,k}}^2\!=\!\sum_{u=1}^{U}\!\sum_{b=1}^{B}\!\sum_{a=1}^{A}\!\hat{w}\!\left(\!\frac{\alpha_{u}t_a^2}{\left(\!\beta_{u}\beta_{b}\!\right)^2}\!+\!\left(\frac{\alpha_{u}t_a}{\!\beta_{u}\beta_{b}}\!-\!\EE[c_{m,k}]\!\right)^2\right)\!\notag,\\ \text{where}\quad\EE[c_{m,k}] = \sum_{u=1}^{U} \sum_{b=1}^{B} \sum_{a=1}^{A} \hat{w} \cdot \frac{\alpha_{u} t_a}{\beta_{u} \beta_{b}}.
\end{align}
\vspace{-8mm}
\subsection{Mutual Coupling}
\ac{MC} occurs when closely spaced antennas interfere with each other's radiation patterns, which in turn impacts the overall channel characteristics between the transmitter and receiver. Accurate \ac{MC} modeling is essential at the \ac{BS}, as neglecting these effects can lead to significant performance degradation, even more so than overlooking spatial correlation~\cite{KolomvakisExploiting}. In this work, we account for \ac{MC} at the \ac{BS} side.\footnote{Recent studies~\cite{Zheng2024Mutual} have shown that the \ac{MC} effect between \ac{ris} unit cells is significantly increased with higher amplification factors of active \ac{ris} or tightly integrated surfaces. In this work, we adopt a passive \ac{ris} and, therefore, neglect the \ac{MC} effect on the \ac{ris} side. Future extensions of this work will account for \ac{ris} \ac{MC}, where the \ac{ris} reflection matrix structure will differ~\cite{Zheng2024Mutual}.}
The coupling matrix transforms the original coupling-free spatial correlation matrix, \(\Rm_{\Bpow \Rpow}\) in \eqref{H_BR}, into the effective spatial correlation matrix~\cite{Ivrlavc2010Toward}
\begin{equation}
\Rm_{\Bpow \Rpow}^{\mathrm{MC}} = \mathbf{M}^{1/2} \Rm_{\Bpow \Rpow} \mathbf{M}^{1/2},
 \quad \text{where} \ \, \mathbf{M} = (\mathbf{Z} + r_d \mathbf{I})^{-1},
\end{equation}
where each antenna element has a dissipation resistance \( r_d\! >\!0 \), accounting for internal losses and becomes significant with small spacing \cite{D2024Holographic}, and \( \mathbf{Z} \) is the mutual impedance matrix, capturing the electromagnetic coupling across elements. For a half-wavelength dipole, $r_d$ is provided in~\cite[Example 2.13]{Balanis2016Antenna}, and closed-form expressions for dipoles in a parallel-in-echelon configuration are available~\cite[eq. (8.73a-b)]{Balanis2016Antenna}.

\section{Channel Estimation}
\label{sec:ch_est}
The \ac{ls} method is widely used for channel estimation, as detailed in \cite{Jensen2020Optimal}. For \(\cv\) to be uniquely estimated, \(\Qm\) must be full rank, requiring \(T_\ppow \geq K\). The \ac{ls} estimator is given by
\begin{equation}
\hat{\cv}_{\mathrm{LS}} = \frac{1}{\sqrt{\rho}}(\Qm^\HH \Qm)^{-1}\Qm^\HH\yv,
\end{equation}
where \(\rho\) is the transmitted pilot power, the training \ac{snr} is defined as \(\gamma = \frac{\rho}{\sigma^2}\). The error covariance matrix from DFT-based training is given in \cite[eq. (15)]{Pan2022Overview} as
\begin{equation}
\Rm_{e_\lsidx}= \frac{1}{\gamma T_\ppow} \mathbf{I}_{M K}.\label{R_{e_{LS}}}
\end{equation}

The \ac{lmmse} estimator minimizes the \ac{MSE} and relies on the second-order statistics of the channel and noise, which can further be expressed as 
\begin{equation}
\hat{\cv}_{\mathrm{LMMSE}} = \sqrt{\rho} \Rm_{\cpow\cpow}\Qm^\HH(\rho \Qm\Rm_{\cpow\cpow}\Qm^\HH + \sigma^2 \mathbf{I}_{M T_\ppow})^{-1}\yv,
\end{equation}
where \(\Rm_{\cpow\cpow}\) is the channel covariance matrix structured as \cite{Pan2022Overview}
\begin{equation}
\Rm_{\cpow\cpow} = (\Rm_{\Rpow \Upow} \odot \Rm_{\Rpow\Bpow}) \otimes \Rm_{\Bpow\Rpow}.
\end{equation}
The \ac{lmmse} error covariance matrix using the DFT-based training phase shift matrix is \cite[eq. (22)]{Pan2022Overview} 
\begin{equation}
\Rm_{e_\lmmseidx} = \left( \Rm_{\cpow\cpow}^{-1} + \gamma T_\ppow \mathbf{I}_{M K} \right)^{-1}. \label{R_{e_{LMMSE}}}
\end{equation}
\section{Numerical Results}
\label{sec:res}

\begin{figure*}[htb!]
 \centering
 \subfloat[Effect of spatial correlation in different fields.]{\label{fig:image2} \includegraphics[width=0.32\linewidth]{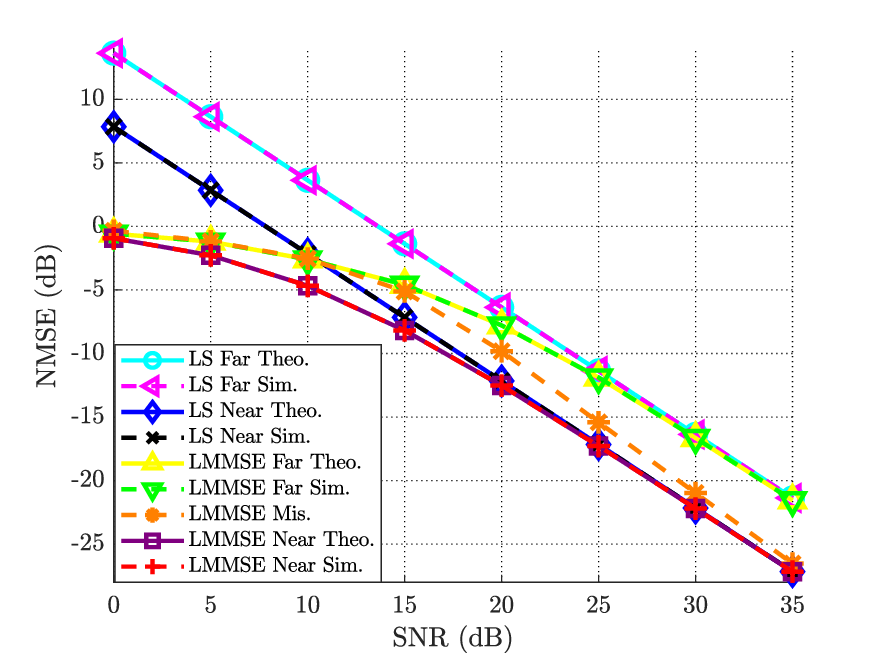}}%
 \hfill
 \subfloat[Impact of considering \ac{MC} in near fields.]{\label{fig:image3} \includegraphics[width=0.32\linewidth]{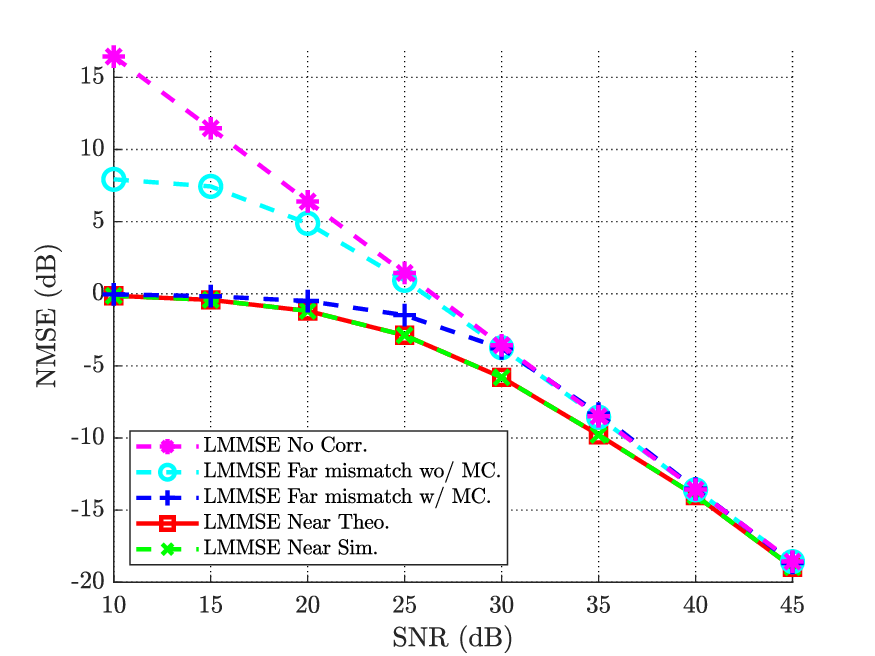}}%
 \hfill
 \subfloat[Effect of changing MG $\alpha$ parameter.]{\label{fig:image4} \includegraphics[width=0.32\linewidth]{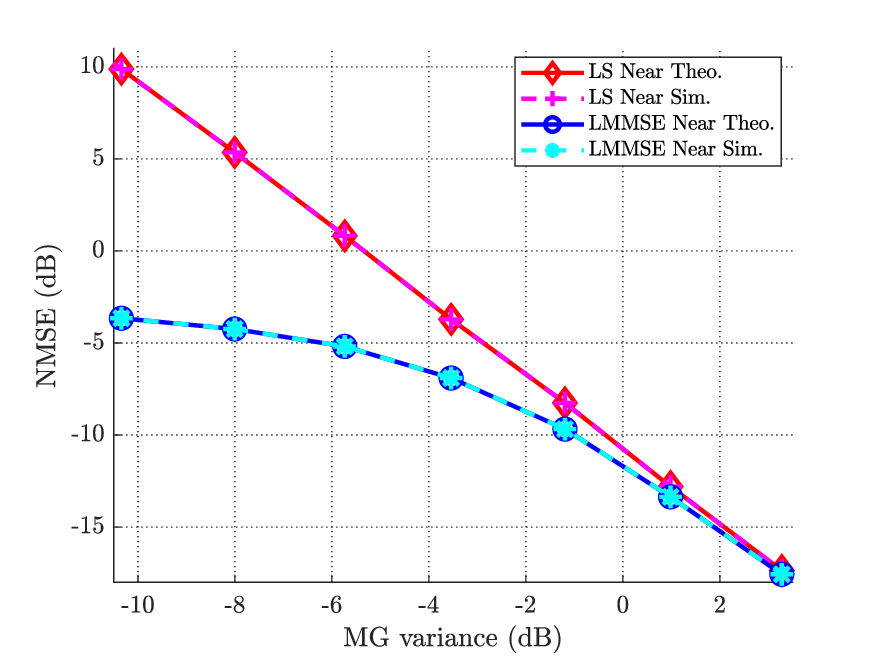}}%
 \vspace{-1mm}
 \caption{Comparison of channel estimation accuracy for different scenarios expressed by NMSE versus: (a), (b) SNR, and (c) MG variance.}
 \label{fig:mainresults}
 \vspace{-4mm}
\end{figure*}
This section evaluates the estimation accuracy of different estimators by utilizing the \ac{NMSE}, where \(\text{NMSE} = \text{Tr}(\mathbf{R}_e)/\text{Tr}(\mathbf{R}_{\cpow\cpow})\). We plot the theoretical curves using \eqref{R_{e_{LS}}} and \eqref{R_{e_{LMMSE}}}, denoted as `$\text{Theo}$', along with the numerical results obtained using Monte Carlo simulations with 1000 trials, labeled as `$\text{Sim}$', in Figs.~\ref{fig:image2}, \ref{fig:image3}, and \ref{fig:image4}. Simulation parameters include \( f_c = 142 \) gigahertz (GHz), system bandwidth \( B = 4 \) GHz, \( \delta_\Bpow = \delta_\Rpow = \lambda / 2 \), and $L=L'=3$, where \( \varphi_{\Upow,l},  \varphi_{\Bpow,l}, \text{ and } \varphi_{\Rpow,l} \) were randomly chosen within \([-\pi/2, \pi/2)\). Similarly, the distances \( D_l \) are set within the radiative near field with a scatterer radius \( r_l = 1.8 \) m for both channels, while in the far-field \( r_l = 4 \) m. Measurement-based \ac{MG} parameters are utilized for a realistic scenario~\cite{Papasotiriou2023Outdoor}. 

In Fig.~\ref{fig:image2}, we evaluate the impact of spatial correlation at the \ac{ris}, setting the \ac{ris} and \ac{BS} element counts to \( K = 128 \) and \( M = 128 \), respectively. We compare estimation accuracy between the near-field and far-field scenarios. Results show that near-field modeling outperforms far-field modeling due to stronger correlation, improving estimation accuracy by up to $\unit[3]{dB}$ over far-field models, at $\unit[10]{dB}$ \ac{snr}. This improvement can be explained by the distinctive nature of near-field signal propagation, where spherical wavefronts prevail, resulting in stronger spatial correlation and more concentrated energy distribution. We also assess the impact of assuming a planar wavefront while the ground truth is near-field by using a mismatch model, by applying the far-field formulation, to demonstrate that this assumption results in an estimation error of approximately 3.5 dB at an \ac{snr} of 15 dB.

We illustrate in Fig.~\ref{fig:image3} the effects of \ac{MC} and spatial correlation at the \ac{BS}. At $\unit[20]{dB}$ \ac{snr}, the results indicate that neglecting both \ac{MC} and spatial correlation leads to a performance degradation of approximately 8 dB. When \ac{MC} is ignored and far-field assumptions are applied, the mismatch penalty is around 6 dB. However, when far-field assumptions are coupled with consideration of \ac{MC}, the error is reduced to 1 dB, highlighting the critical importance of including \ac{MC} in the \ac{BS} for accurate system modeling. These findings are tied to electromagnetic properties, where neglecting \ac{MC} causes a significant mismatch in the received signal power.
Finally, Fig.~\ref{fig:image4} illustrates the effect of the shape parameter \(\alpha\) of the \ac{MG} distribution on channel variance and estimator performance for fixed $\rho$. The parameter \(\alpha\) reflects the degree of fading severity, with larger values indicating less fading. As \(\alpha\) increases, channel variance increases, suggesting a greater contribution from non-\ac{los} components or scatterers.

\section{Conclusions}
\label{sec:conc}
This paper, we study a \ac{thz}-band \ac{ris}-aided communication system using the \ac{MG} distribution to model small-scale fading in outdoor scenarios, while incorporating near-field spatial correlation for large arrays at the \ac{BS} and \ac{ris}. We also account for \ac{MC} at the \ac{BS} and key \ac{thz}-specific effects, such as molecular absorption loss. Based on this model, we evaluate the performance of linear channel estimators. Results show that near-field spatial correlation improves estimation accuracy over far-field models, especially with ultra-massive arrays. Moreover, ignoring \ac{MC} at the \ac{BS} leads to notable estimation errors and performance loss.

\bibliography{IEEEabrv,ref}
\bibliographystyle{IEEEtran}
\end{document}